# Structural Reconstruction in Lead-free Two-dimensional Tin Iodide Perovskites Leading to High Quantum Yield Emission

*Sushant Ghimire,[a] Kevin Oldenburg,[b] Stephan Bartling,[c] Rostyslav Lesyuk,[a] Christian Klinke*[a,b,d]*

[a] Institute of Physics, University of Rostock, Albert-Einstein-Strasse 23, 18059 Rostock, Germany

[b] Center for Interdisciplinary Electron Microscopy (ELMI-MV), Department "Life, Light & Matter", University of Rostock, Albert-Einstein-Strasse 25, 18059 Rostock, Germany

[c] Leibniz Institute of Catalysis (LIKAT), Albert-Einstein-Strasse 29a, 18059 Rostock, Germany

[d] Department of Chemistry, Swansea University, Swansea SA2 8PP, United Kingdom

**Corresponding Author**

*Email: christian.klinke@uni-rostock.de

ABSTRACT: We report a structural reconstruction-induced high photoluminescence quantum yield of 25% in colloidal two-dimensional tin iodide nanosheets that are synthesized by a hot-injection method. The as-synthesized red-colored nanosheets of octylammonium tin iodide perovskites at room temperature transform to white hexagonal nanosheets upon washing or exposure to light. This structural change increases the bandgap from 2.0 eV to 3.0 eV, inducing a large Stokes shift and a broadband emission. Further, a long photoluminescence lifetime of about 1 µs is measured for the nanosheets. Such long-lived broad and intense photoluminescence with large Stokes shift is anticipated to originate from tin iodide clusters that are formed during the structural reconstruction. Stereoactive $5s^2$ lone pair of tin (II) ions perturbs the excited state geometry of the white hexagonal nanosheets and facilitates the formation of self-trapped excitons. Such broadband and intensely emitting metal halide nanosheets are promising for white light-emitting diodes.

**TOC GRAPHICS**

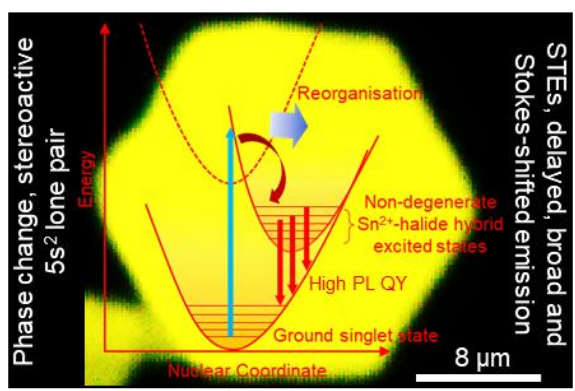

Three-dimensional (3D) lead halide perovskites have shown promising results in solar cells,[1,2] light-emitting diodes (LEDs),[3] lasers,[4] and photodetectors.[5] However, Goldsmith tolerance factor and octahedral factor restrict the ABX$_3$ structure of 3D perovskites to a narrow choice of A-site cations to methylammonium (CH$_3$NH$_3^+$, MA), formamidinium (HC(NH$_2$)$_2^+$, FA), and Cs, B-site cations to Pb, Sn, and Ge, and X anions to halides (Cl$^-$, Br$^-$ and I$^-$).[6,7] The ionic nature of the crystal lattice makes 3D halide perovskites highly unstable in moisture and under intense photoirradiation.[8–10] On the other hand, two-dimensional (2D) halide perovskites, which are formed by cutting the ABX$_3$ structure at the (100), (110), or (111) crystalline plane,[11] overcome the limitations imposed by their 3D analogs. They can accommodate large A-site cations whose size lay beyond the tolerance limit.[12] The medium to bulky mono- or diammonium organic spacers sandwich the inorganic layers in between and keep them out of reach of water and moisture which improves their stability.[13,14] The photostability is ensured by the suppressed ion migration along the stacking plane.[15] Such stable 2D halide perovskites are promising for high-performance photovoltaics and photocatalysis.[11,16] Recently, a stable 2D halide perovskite solar cell with a record high efficiency exceeding 21% was reported by Shao et al.,[17] demonstrating their promising nature under practical environmental conditions.

A 2D halide perovskite has a quantum well structure where the inorganic layer acts as well and the organic spacer as a barrier. This results in interesting optoelectronic processes such as strong quantum and dielectric confinement,[11] optical anisotropy,[18] exciton dissociation at the crystal edge,[19] self-trapped excitons (STEs),[20] and prolonged hot-carrier cooling.[21] Further, the organic spacers can be functionalized to tune the electronic, optical, and charge-transport properties of 2D halide perovskites.[11,22] While strongly bound excitons in low-dimensional halide perovskites result in fast and narrow emission originating from their band-edge radiative relaxation, largely Stokes shifted broad and delayed emission is also observed in them which is attributed to defects and STEs. [20,23–33] Broadband emission from low-dimensional perovskites and metal halide layered structures are important for producing white LEDs.[34–36] STEs are formed when exciton creates a potential well by deforming the lattice locally and getting trapped into it. While emissive defect states in perovskites are extrinsic and are accessible with below-bandgap excitation,[31,33,37] STEs are intrinsically formed and are not accessible with below-bandgap excitation.[11,24] Emission from STEs is temperature-dependent where a sufficiently low temperature is favorable for excitons to access the self-trapping states.[20,23,34] At room temperature or higher, a thermal detrapping of STEs and quenching of broadband emission may take place. Therefore, producing broadband white light emission from halide perovskites at room temperature is challenging. Recently, Kanatzidis and coworkers demonstrated a huge structural diversity among the room temperature white light-emitting one-dimensional (1D), 2D, and 3D perovskites and perovskitoids.[36] Zhou et al. demonstrated an STE-based broadband yellow light emission with high photoluminescence (PL) quantum yield (QY) of 85% from the single crystals of 0D $(C_4N_2H_{14}Br)_4SnBr_xI_{6-x}$ perovskite.[28] Nevertheless, in the case of tin- and germanium-based low-dimensional perovskites, structural distortion in the excited state is mainly governed by the stereoactivity of $ns^2$ lone-pair electrons which ultimately affect their excitonic properties.[26,27,38–40]

On top of these interesting optical properties, monolayered and colloidal 2D halide perovskites suffer from nonradiative Auger processes induced by strong carrier-carrier interaction and surface defects which greatly compromise their PL QY.[41] Further, the environmental toxicity of Pb has raised questions on the wide-range applications of lead-based halide perovskites despite their superior optoelectronic properties and device performance.[42] Lead-free perovskites containing Sn or Ge, on the other hand, bear comparatively low PL QY and poor performance in solar cells and LEDs which is due to the increased nonradiative recombination caused by the abundant vacancy defects that are formed through oxidation of $Sn^{2+}$ to $Sn^{4+}$ and $Ge^{2+}$ to $Ge^{4+}$.[43,44] Further, such easy oxidation makes tin- or germanium-based halide perovskites intrinsically unstable.[45,46] These limitations demand the preparation of lead-free 2D halide perovskites or perovskite-like materials in their colloidal form which show suppressed nonradiative recombination, high PL QY, and better stability.

In this work, we demonstrate a high PL QY of 25% and a delayed PL of about 1 µs in colloidal 2D tin iodide nanosheets. These nanosheets are obtained either by a direct synthesis using a modified hot-injection technique or by a structural reconstruction of 2D octylammonium tin iodide Ruddlesden-Popper perovskites at room temperature upon washing or exposure to light. In a conventional hot-injection method, the shape of colloidal halide perovskites is controlled by varying the reaction temperature and the chain length of ligands.[47] Further, the reaction is quenched by quickly cooling the reaction mixture in an ice bath. Therefore, it is often difficult to grow micron-size nanosheets by this method. Our method is a modified hot injection in which octylamine and tri-*n*-butylphosphine (TBP) as ligands and 1,2-diiodoethane as iodide source are subsequently injected into a hot solution of tin (II) oleate at 80 °C, and the whole mixture is allowed to cool slowly (~1 °C/min) to room temperature. The 2D nanosheets with a lateral size of several microns precipitate during cooling. The use of alkyl halides as the source of halide ions in the synthesis of 2D perovskites was demonstrated by us in our previous report.[48] Figure 1a shows the scheme for the synthesis of 2D octylammonium tin iodide nanosheets. The slow precipitation during cooling is anticipated to take place by the slow release of iodine from alkyl iodide which provides enough time for nuclei to grow into large sheets. Also, a large amount of oleic acid in the reaction mixture directs the templated growth of 2D structures.[49] Interestingly, red-colored nanosheets were formed when tin oleate was dried under a high vacuum of about $10^{-2}$ mbar, whereas white-colored samples were obtained when the vacuum was 1 mbar (Figure 1a). Under the weak vacuum condition, acetic acid formed as a by-product during the reaction of tin acetate with oleic acid is not completely removed from the reaction mixture. Hence, the presence of a trace amount of such polar species in the reaction mixture results in the structural reconstruction of red-colored octylammonium tin iodide nanosheets. The structural reconstruction of nanosheets also takes place when they are washed with solvents such as hexane or toluene and/or exposed to light.

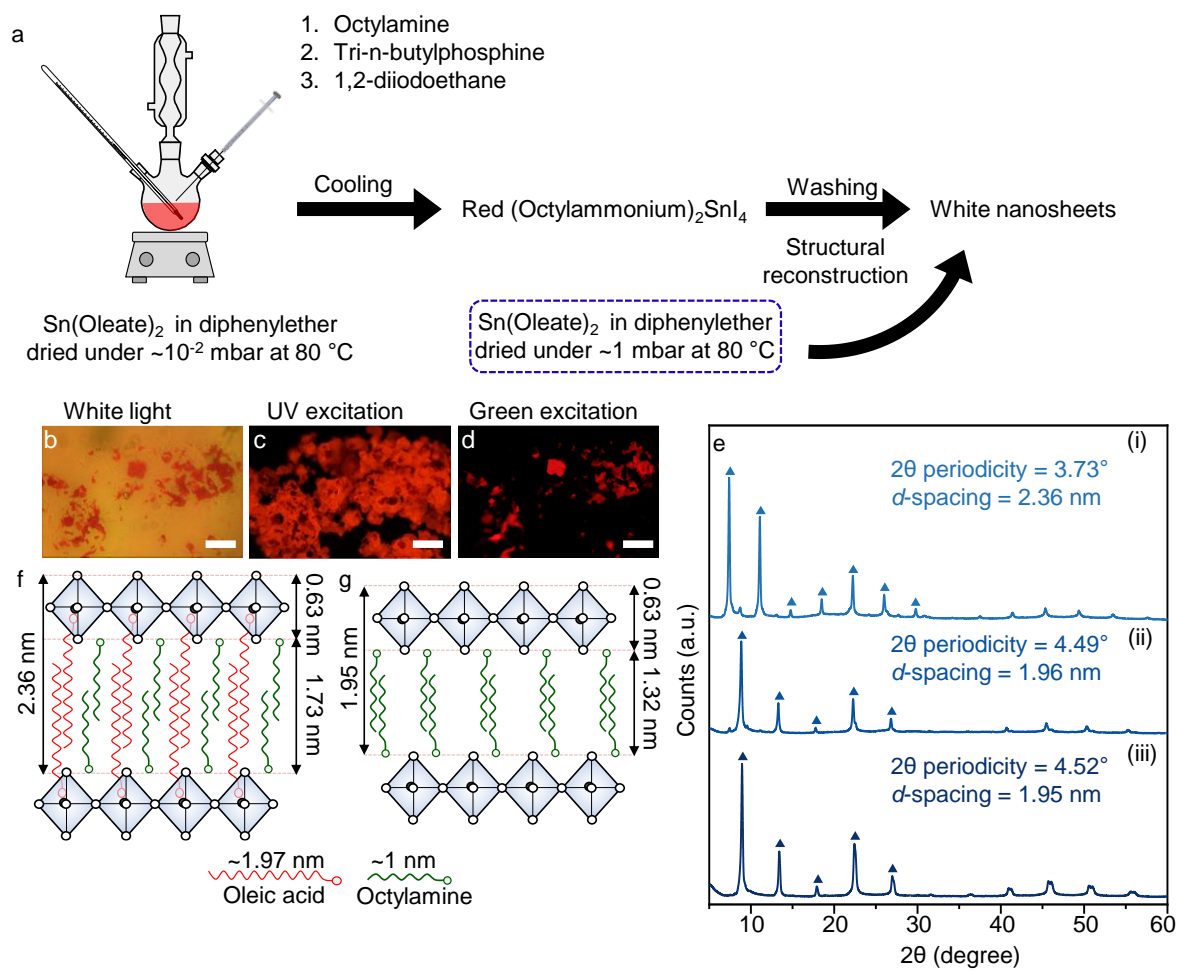

**Figure 1.** Synthesis and characterization of 2D octylammonium tin iodide Ruddlesden-Popper perovskites. (a) Scheme showing the synthesis steps. (b-d) Fluorescence optical microscopy images under (b) room light, (c) UV excitation, and (d) green excitation. Scale bars are 50 µm. (e) Powder XRD patterns of (i) as-synthesized red-colored octylammonium tin iodide perovskite nanosheets, (ii) white hexagonal nanosheets obtained from the repeated washing of the red-colored nanosheets, and (iii) as-synthesized white hexagonal nanosheets. The triangles indicate the peaks for which the periodicity and *d*-spacing are calculated. (f, g) Schematic representation of the (f) red-colored octylammonium tin iodide nanosheets and (g) white hexagonal nanosheets stacking with the *d*-spacing calculated from (e).

We characterized the samples using fluorescence optical microscopy (Figure 1b-d), powder X-ray diffraction (XRD, Figure 1e), scanning electron microscopy (SEM, Figure S1), and scanning transmission electron microscopy (STEM, Figure S1). As shown in Figure 1b, optical microscopy of a sample reveals a red-colored and transparent appearance of the nanosheets under room light. The transparent nanosheets are hexagonal, and they appear white when observed with naked eyes. We referred to them as (structurally reconstructed) white hexagonal nanosheets. While the red-colored octylammonium tin iodide and the white hexagonal nanosheets are both luminescent under UV light (Figure 1c), only the red-colored nanosheets emit during green excitation (Figure 1d), indicating significantly different bandgaps and different optical properties in them. Further, the powder XRD patterns of their films (Figure 1e) show periodically spaced diffraction peaks below 30° 2θ angle which is attributed to the oriented growth and periodic stacking of 2D layered structure. We confirm from powder XRD that the as-synthesized white hexagonal nanosheets and those obtained by repeated washing of red-colored octylammonium tin iodide monolayers with hexane are both derived from the structural reconstruction of the parent perovskites. Here, the powder XRD of the corresponding film of hexane-washed octylammonium tin iodide nanosheets [Figure 1f (ii)] matches well with that of the as-synthesized white hexagonal nanosheets [Figure 1f (iii)]. We anticipate that either the presence of polar species in the reaction mixture or the repeated washing of the nanosheets by solvents results in the removal of surface ligands, prompting the structural reconstruction and transformation. Besides, the diffraction peaks are shifted to lower 2θ angles for the red-colored nanosheets [Figure 1f (i)] when compared to the white hexagonal nanosheets [Figure 1f (ii) and (iii)] which indicates a smaller stacking distance in the latter case. We calculated *d*-spacing using 2θ periodicity of 3.73° and 4.52° for the red-colored and the white hexagonal nanosheets, respectively. The red-colored octylammonium tin iodide stacks at a distance of 2.36 nm. Considering the size of a tin iodide octahedron at 0.63 nm,[50] the space available for the protonated octylamine is 1.73 nm (Figure 1f), which is larger than the size of a single amine ligand (~1 nm in length) but less than that of two. This indicates the octylammonium ligands should stay interdigitated with those of the adjacent layer. Interestingly, the calculated *d*-spacing can accommodate the oleic acid ligand (length is ~1.97 nm) if an additional ligand-binding mode of deprotonated oleic acid with $Sn^{2+}$ is considered. In the case of $APbX_3$ perovskites, the binding of oleate ions to $Pb^{2+}$ on the surface of the nanocrystals is reported.[51,52] The inclusion of larger oleic acid ligands in the interlayer spacing accounts for the larger stacking distance in red-colored octylammonium tin iodide nanosheets. On the other hand, the *d*-spacing decreases to 1.95 nm in white hexagonal nanosheets. As shown schematically in Figure 1g, only 1.32 nm interlayer space is available which can accommodate the octylammonium ligand but not the oleic acid. Consequently, this decreases the stacking distance in the white hexagonal nanosheets. In other words, structural reconstruction follows the expulsion of oleic acid ligands from the interlayer spacing and the rearrangement of the lattice with octylamine.

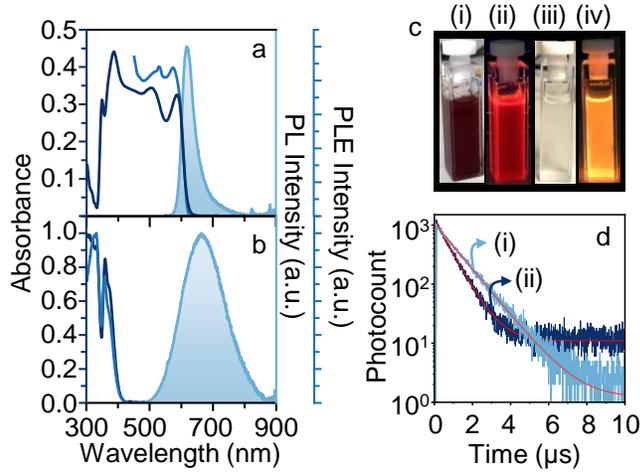

**Figure 2.** Optical properties of 2D octylammonium tin iodide Ruddlesden-Popper perovskite nanosheets. (a,b) Absorption, PL excitation, and PL spectra of octylammonium tin iodide (a) red-colored nanosheet and (b) white hexagonal nanosheet suspensions. (c) Cuvette images of octylammoium tin iodide (i, ii) red-colored nanosheets and (iii, iv) white hexagonal nanosheets under (i, iii) room light and (ii, iv) 365 nm UV light. (d) Time-resolved PL decay profiles of octylammoium tin iodide (i) red-colored nanosheets and (ii) white hexagonal nanosheets. The samples were excited using a 375 nm picosecond laser.

We studied the optical properties of the samples using steady-state and time-resolved PL and UV-Vis absorption spectroscopy. Figures 2a and 2b show the absorption, PL, and PL excitation spectra of red-colored octylammonium tin iodide nanosheets and as-synthesized white hexagonal nanosheets, respectively. Here, the red-colored nanosheets show an absorption onset at 640 nm and the excitonic peak at 586 nm. The calculated bandgap of the sample is 1.99 eV. Under green (530 nm) excitation, the sample shows emission at 618 nm with a full width at half maximum (FWHM) of 45 nm. A small Stokes shift of 32 nm and well-matched PL excitation and absorption spectra indicates that the emission in the case of red-colored nanosheets is from the band edge. On the other hand, the as-synthesized white hexagonal nanosheets show the absorption onset at 423 nm and distinct dual excitonic bands at 360 nm and 375 nm. The calculated bandgap of the sample is 3.12 eV. Here, although the PL excitation spectrum is identical to the absorption spectrum of the sample, a broad emission (FWHM=138 nm) with the maxima at 670 nm and a large Stokes shift of 295 nm is observed. These results suggest that the broadband emission in the white hexagonal nanosheets does not originate from the band edge relaxation of the free excitons. Also, there should not be any defect-related emissive mid-gap states available for exciton trapping since no additional absorption or PL excitation feature was observed below the bandgap. Rather, the origin of such broad and largely Stokes shifted emission should have a connection to the STEs which are facilitated in our case by the structural reconstruction.

We also consider the possibility of octahedral distortion of inorganic layers in our nanosheets which can influence the formation of STEs and optical properties. In general, the octahedral tilting in halide perovskites decreases the metal and halide orbital overlap and increases the optical bandgap.[11] In 2D halide perovskites, out-of-plane distortion depends upon the composition of organic spacers, and it favors the formation of STEs by enhancing the optical deformation potential strength.[11,53] For comparison, we synthesized red-colored 2D tetradecylammonium tin iodide perovskites under similar conditions [Figure S2 (a, b)]. Contrary to the octylammonium tin iodide nanosheets, these 2D perovskites with longer alkylammonium chains did not change to white hexagonal crystals when washed or exposed to light. Rather, they maintained a narrow emission (FWHM=36 nm) with a small Stokes shift [35 nm, Figure S2 (c)] and were stable over time [Figure S2 (d)]. This indicates the different extent of stress exerted on the inorganic slabs that are hydrogen-bonded to the short-chain and the long-chain ligands which in turn regulates the structural reconstruction and octahedral distortion in them.[54] These results are indirect evidence showing a greater tendency of octahedral distortion in octylammonium tin iodide nanosheets due to larger stress developed on the inorganic slab owing to their short carbon chain alkylammonium ligands. On the other hand, the presence of the long carbon chain ligands in 2D tetradecylammonium tin iodide perovskites has a lesser tendency of octahedral distortion due to the minimization of the stress on the octahedral inorganic layer. Figure 2c shows the cuvette images of the colloidal suspensions of red-colored octylammonium tin iodide and white hexagonal nanosheets under room light and UV light. We measured the absolute PL QY of these samples (Supporting Information and Figure S3), revealing it to be below 1% for the red-colored nanosheets and 25% for the white hexagonal nanosheets. Such a huge jump in the PL QY is induced by the structural reconstruction. We calculated the average PL lifetime of the samples by fitting the decay profiles with a bi-exponential decay function (Supporting Information and Table S1). The white hexagonal nanosheet samples show an average PL lifetime of 734 ns [Figure 2d(i)]. Such a delayed PL is anticipated to result from the self-trapping of excitons.[55,56] Nevertheless, the average PL lifetime of the red-colored octylammonium tin iodide nanosheets [Figure 2d(ii)] is longer (1.12 µs) than that of the as-synthesized white hexagonal nanosheets, indicating that the sample at its early phase of structural reconstruction contains the domains of both the small bandgap red-colored and the large bandgap white hexagonal nanosheets. In such a system with different bandgap domains, there is a possibility of energy transfer from the larger bandgap to the smaller bandgap, resulting in delayed emission.[11,48] However, once the red-colored nanosheets are completely transformed, as in the case of as-synthesized white hexagonal nanosheets, the energy transfer is poor since there is no overlap between the absorption and the emission spectra. In such a case, PL is governed by STEs.

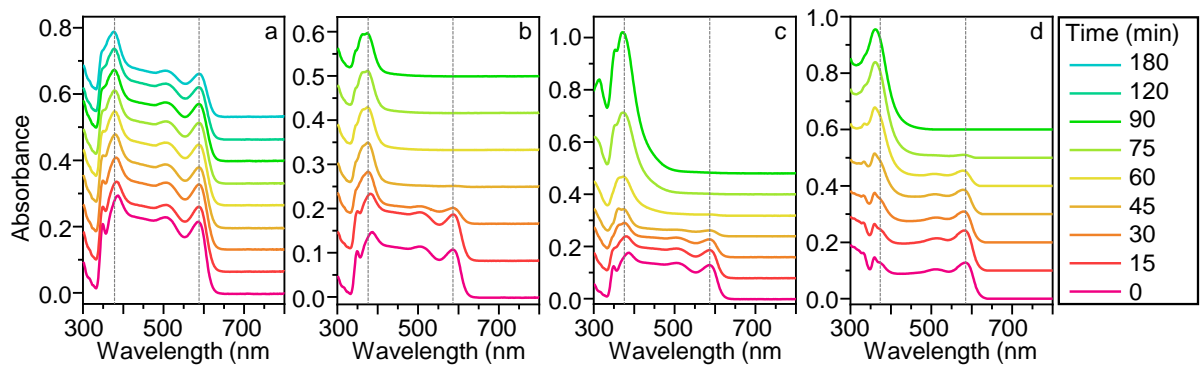

**Figure 3.** Structural reconstruction in 2D octylammonium tin iodide Ruddlesden-Popper perovskite nanosheets. Absorption spectra of (a-c) red-colored octylammonium tin iodide nanosheet suspension (a) in dark, (b) under room light, and (c) under 365 nm UV light, and (d) hexane-washed red-colored octylammonium tin iodide nanosheet suspension stored in dark. All spectra are recorded for 90 to 180 min at a time interval of 15 min.

We monitored the structural reconstruction and transformation of red-colored octylammonium tin iodide nanosheets to white hexagonal nanosheets by continuously recording the UV-Vis absorption spectra in the dark, room light, UV light, and after washing with hexane (Figure 3). The observed transformation of red-colored nanosheets to white hexagonal ones was irreversible. As shown in Figure 3a, the sample shows an excitonic peak at 586 nm in the dark which remains intact over 180 min. However, a second absorption peak at 375 nm becomes prominent, indicating the structural reconstruction and transformation. This structural change in the dark is slow and incomplete as the red-colored nanosheets coexist. On the other hand, under room light (Figure 3b) and UV light (Figure 3c), the red-colored nanosheets change to white more quickly and completely as indicated by the complete disappearance of the excitonic peak at 586 nm. Also, the red-colored nanosheets under UV irradiation start to decompose after 60 min and show a relatively broad absorption onset at 550 nm. Lanzetta et al. have shown the release of molecular iodine during the degradation of tin iodide perovskites which was indicated by a broad absorption at 500 nm.[57] Apart from light, the structural reconstruction was observed when the red-colored nanosheets were washed with hexane as shown in Figure 3d. Here, immediately after solvent washing, the sample showed the absorption features of the white hexagonal nanosheets in addition to that of the red-colored ones. Over time, the absorption from the red-colored nanosheets disappeared and only the absorption features from the white hexagonal nanosheets remained, indicating a complete structural reconstruction and transformation. Nevertheless, we do not rule out the degradation of the samples over time. Figure S4 shows the cuvette images of red-colored octylammonium tin iodide nanosheet suspensions under different conditions indicating stability, structural transformation, and degradation. The sample in dark remains stable over time which is indicated by the red color of the suspension that emits under the UV excitation. On the other hand, the samples exposed to room light and UV light degrade over time which is indicated by their yellowish or brownish coloration in the cuvette at room light. These degraded samples are non-luminescent when photoexcited. One interesting observation of our work is the structural transformation and the degradation of 2D halide perovskites when exposed to UV-visible light and washed with solvent. This is contrary to the widely reported results that discuss the intrinsic stability of 2D halide perovskites contributed by the bulky organic ligands.[14,15,17,58] A recent report by Kamat and coworkers shows the photo-induced ion migration and phase segregation in 2D halide perovskites which depends on the type of organic cation used.[59] Monolayered 2D halide perovskites with aliphatic alkylammonium ligands are more prone to photo-induced phase segregation due to a weaker van der Waals interaction among the stacking layers than the one with the aromatic functional group which bears much stronger π-π interaction. The ion migration in halide perovskites also depends on temperature[15] and the composition of halide ions.[60] Such ion migration leads to the degradation of 2D halide perovskites.[61] We anticipate that ion migration can trigger the rearrangement of the crystal lattice and degradation of samples in our case too.

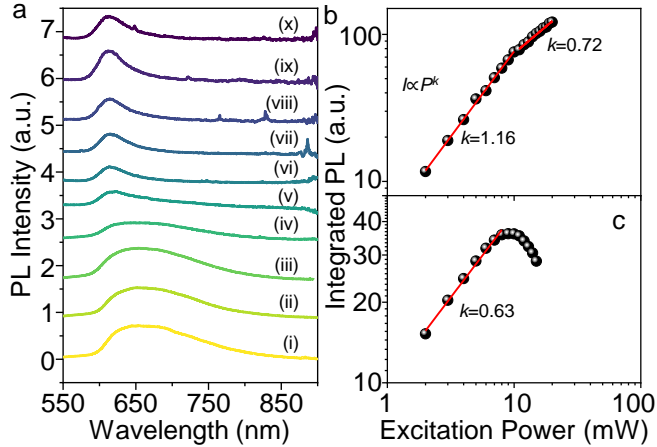

**Figure 4.** Narrow excitonic and broadband STE-based emissions in 2D octylammonium tin iodide nanosheets. (a) PL spectra of red-colored octylammonium tin iodide nanosheet suspension at different excitation wavelengths: (i) 350 nm, (ii) 370 nm, (iii) 390 nm, (iv) 410 nm, (v) 430 nm, (vi) 450 nm, (vii) 470 nm, (viii) 490 nm, (ix) 510 nm, (x) 530 nm. (b, c) Integrated PL intensity versus excitation power plots for (b) red-colored octylammonium tin iodide and (c) white hexagonal nanosheets.

We distinguished the band edge narrow excitonic emission from the broadband emission by recording the excitation wavelength-dependent PL spectra of the red-colored octylammonium tin iodide nanosheets at their early stage of structural reconstruction. As shown in Figure 4a, the sample shows a broad emission spawning a range from 575 to 840 nm when excited at 350 nm. On systematically increasing the excitation wavelength from 350 nm to 530 nm, the emission gradually blue shifts and becomes narrow, centering around 614 nm. This shows emissions from two different red-colored and white hexagonal phases with distinct bandgaps, which are accessible with different excitation energies. On the other hand, no such change in the position and shape of PL spectra with excitation energy was observed in the case of white hexagonal nanosheets (Figure S5). The absence of emission below the bandgap excitation in the case of white hexagonal nanosheets indicates that there are no localized mid-gap defect-related emissive energy states. To clarify further, we studied the excitation intensity-dependent steady-state PL of these samples by exciting them above their bandgaps using continuous-wave lasers. We excited the red-colored nanosheet suspension at 410 nm and the white hexagonal nanosheet suspension at 375 nm, continuously by varying the intensity in the range of 2 to 20 mW. As shown in Figure 4b, the power-law fitting ($I \propto P^k$, where $I$ is integrated PL intensity, $P$ is excitation power, and $k$ is the power-law exponent) of the log-log plot of integrated PL intensity versus excitation power in the case of red-colored octylammonium tin iodide nanosheets estimates $k$ at 1.16 below and 0.72 above 10 mW. On the other hand, in the case of white hexagonal nanosheets, the value of $k$ is 0.63 below 8 mW, and the sample degrades above this power which is observed as the continuous decrease in the integrated PL intensity. Yin et al. assigned the origin of broadband emission in 2D (PEA)$_2$PbI$_4$ perovskites to the emissive iodide vacancy defects.[31] They showed that the PL intensity varies linearly with the excitation power for free excitons with $k$ equals 1.28. Alternatively, for the broadband emission, $k$ equals 0.96 which indicates defect emission in (PEA)$_2$PbI$_4$. Solely based on the value of $k<1$, the broadband emission in our samples cannot be assigned to the emissive atomic defects. This is because the results from the PL excitation spectra and excitation energy-dependent emission spectra of these samples do not support the presence of emissive defects. The observed trend in the power-law behavior of PL in white hexagonal nanosheets can be explained with the model of Schmidt et al.[62] They discussed the power-law behavior of PL of a direct bandgap semiconductor in terms of free and bound excitons for which 1<k<2 and free-to-bound and donor-acceptor pair recombinations for which k<1. In our case, the localization of excitons may happen within the clusters of tin iodide formed during the structural reconstruction, the mechanism of which is discussed in the following paragraph. Therefore, the self-trapping of free excitons to localized ones and their recombination to produce broadband emission can be considered as free-to-bound excitonic transition and recombination. Nevertheless, we do not rule out the presence of nonradiative defects in our samples.

Tin halide perovskites are known to be self-doped with $Sn^{4+}$ which is formed by the *in-situ* oxidation of $Sn^{2+}$ in the sample. This self-doping modifies the optoelectronic properties of tin halide perovskites.[63–65] To understand whether our samples are self-doped or not, we studied the surface chemical composition using X-ray photoelectron spectroscopy (XPS). As shown in Figure 5a, the white hexagonal nanosheets show Sn mainly in its +2 oxidation state.[66] Although this indicates that the white hexagonal nanosheets are stable towards self-doping, the results are not 100% conclusive since the binding energies of $Sn^{2+}$ and $Sn^{4+}$ are very close to each other.[67] Nevertheless, even if $Sn^{4+}$ is present in our samples, it should be in trace amount (<<1%) which is less likely to affect the observed optoelectronic properties. Further, the XPS shows the interaction of oxygen with the surface of 2D octylammonium tin iodide monolayers which is revealed by the presence of peaks corresponding to the $IO^{2-}$ and adsorbed $O^{2-}$ species in the core level spectra of I 3d (Figure 5b) and O 1s (Figure 5d), respectively.[68] Such interaction might have also triggered the structural reconstruction and transformation of our samples. The surface interaction with oxygen leads to the degradation of 2D tin iodide perovskites, which we observed as tarnishing of our samples in their dried powder form to brown or black color when exposed to air.

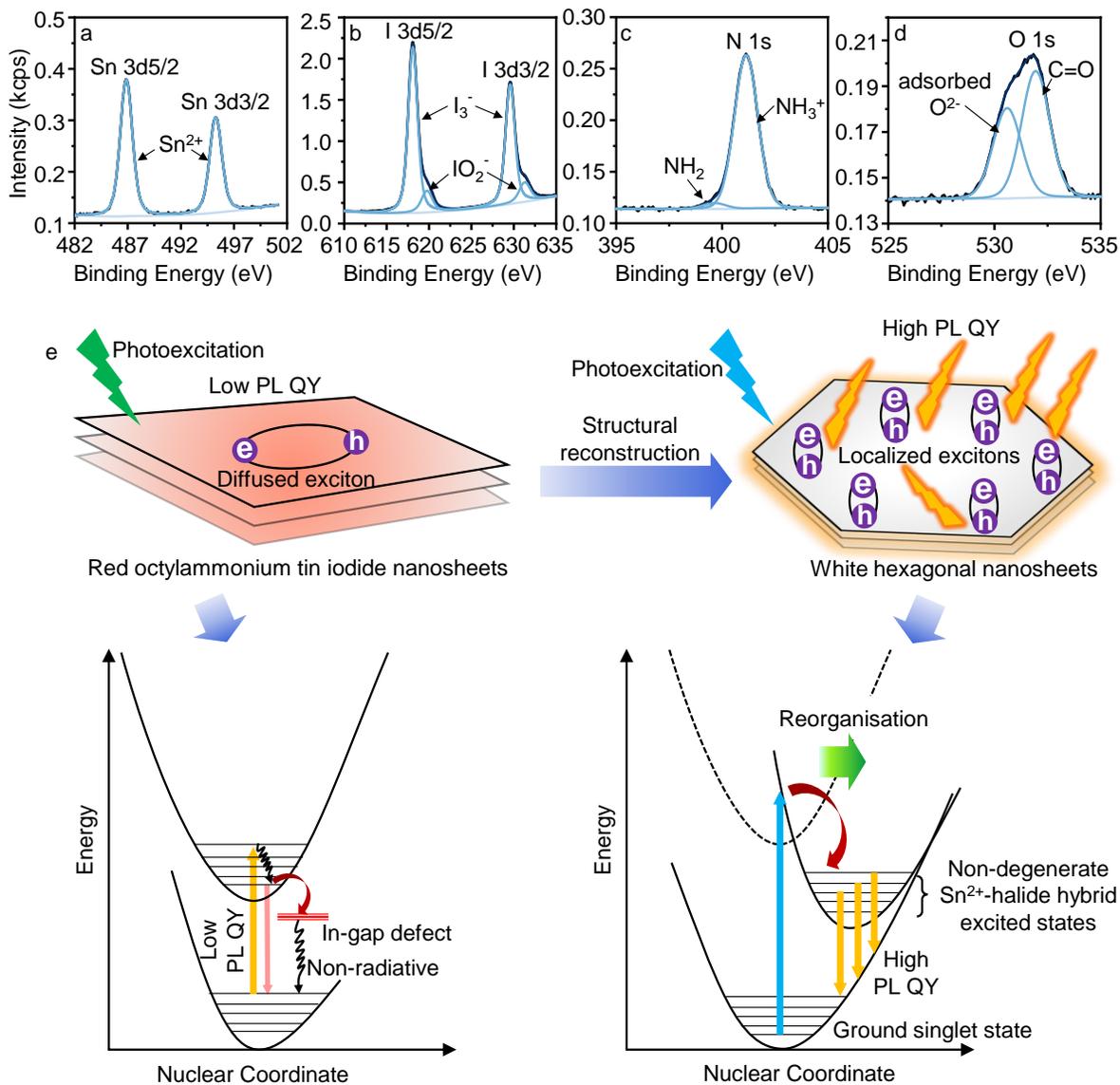

**Figure 5.** Mechanism of broadband emission. (a-d) XPS spectra of octylammonium tin iodide white hexagonal nanosheets that correspond to the core level of (a) Sn 3d, (b) I 3d, (c) N 1s, and (d) O 1s. (e) Scheme showing the mechanism of broadband emission.

We hypothesize that the broadband and high QY emission in white hexagonal nanosheets at room temperature is from the STEs, the origin of which is related to the stereoactive $5s^2$ lone pair electrons in $Sn^{2+}$. As shown schematically in Figure 5e, excitons in red-colored octylammonium tin iodide nanosheets can diffuse laterally, get trapped into the defects, and recombine nonradiatively resulting in low PL QY. In the white hexagonal nanosheets, $5s^2$ lone pairs are anticipated to be strongly expressed through either the formation of 0D tin iodide clusters which are uniformly distributed in the 2D organic-inorganic matrix, or the octahedral distortion of the inorganic layer. The excitons that are localized in the isolated 0D tin iodide clusters recombine radiatively with high PL QY. We discuss the mechanism of STE emission in our case based on the models proposed by Kovalenko and co-workers.[40] The stereoactive lone pair of $Sn^{2+}$ in white hexagonal nanosheets leads to stronger structural distortion and reorganization of the excited state. The difference between the excited- and ground-state geometries, shown schematically by the different positions of the energy minima in the parabola in the lower right panel of Figure 5e, is reflected as the large Stokes shift of emission. The emission transitions occur from different singlet and triplet non-degenerate excited energy states formed by the hybridization of Sn and halide atomic orbitals, while the absorption occurs from the ground singlet state to the higher energy states. The averaging of multiple transitions from the excited to the ground states result in a broadband emission. Also, a large Stokes shift and localized excitons suppress the energy loss to nonemissive defects through poor energy transfer. As a result, the white hexagonal nanosheet samples show high PL QY. Recently, Xu et al. showed a reversible phase transformation between the nonemissive 2D $(PEA)_2SnBr_4$ and highly emissive 0D $[(PEA)_4SnBr_6][(PEA)Br]_2[CCl_2H_2]_2$ single crystals, where the strongly localized excitons in 0D structures emit a broad yellow light with 90% PL QY.[26] Although our samples show lower PL QY when compared to that of the reported low-dimensional tin iodide perovskite single crystals,[27] it is comparable to or improved beyond that of the colloidal ones.[30,50,69] The removal of the surface passivating ligands during the purification step and surface degradation of nanosheets with time induce an additional nonradiative decay channel which is anticipated to diminish the PL QY of our colloidal samples below that of the solid-state single crystals.

In summary, we demonstrate broadband delayed emission with a large Stokes shift and a high PL QY from the colloidal octylammonium tin iodide nanosheets which are formed by the structural reconstruction of their parent 2D perovskites. The structural transformation is irreversible and is triggered by light exposure and solvent washing. Based on the optical studies, we do not consider the defect-related mid-gap energy states responsible for such broadband emission and large Stokes shift. Rather, we formulate the mechanism based on the stereoactive $5s^2$ lone pair in $Sn^{2+}$ and STEs. Such broadband and high PL QY emitting lead-free 2D halide perovskites are attractive for applications in environmentally benign LEDs.

## ASSOCIATED CONTENT

**Supporting Information**

Experimental details, SEM, HRTEM, SAED, absorption, PLE, and PL spectra, XRD, cuvette images

## AUTHOR INFORMATION

**Corresponding author**

Christian Klinke – Institute of Physics and Department "Life, Light & Matter", University of Rostock, Germany and Department of Chemistry, Swansea University, United Kingdom

orcid.org/0000-0001-8558-7389; Email: christian.klinke@uni-rostock.de

**Notes**

The authors declare no competing financial interest.

ACKNOWLEDGMENT

SG acknowledges Alexander von Humboldt Stiftung/Foundation for the postdoctoral research fellowship. We acknowledge the European Regional Development Fund of the European Union for funding the PL spectrometer (GHS-20-0035/P000376218) and X-ray diffractometer (GHS-20-0036/P000379642) and the Deutsche Forschungsgemeinschaft for funding an electron microscope Jeol NeoARM TEM (INST 264/161-1 FUGG), an electron microscope ThermoFisher Talos L120C (INST 264/188-1 FUGG), and supporting the collaborative research center LiMatI (SFB 1477). We acknowledge Prof. Dr. S. Speller and Dr. R. Lange of the University of Rostock for providing/supporting us with the Scanning Electron Microscope.